\documentclass[preprint,showpacs,preprintnumbers,amsmath,amssymb]{revtex4}
\usepackage{epsfig}
\usepackage{graphicx}% Include figure files
\usepackage{dcolumn}% Align table columns on decimal point
\usepackage{bm}% bold math

\newcommand{\ord}{{\cal O}}
\def\beq{\begin{equation}}
\def\eeq{\end{equation}}
\def\eeqn{\end{equation}}
\newcommand\iden{\leavevmode\hbox{\small1\normalsize\kern-.33em1}}

\newcommand{\bea} {\begin{eqnarray}}
\newcommand{\eea} {\end{eqnarray}}

\let\jnfont=\rm
\def\NPB#1,{{\jnfont Nucl.\ Phys.\ B }{\bf #1},}
\def\PLB#1,{{\jnfont Phys.\ Lett.\ B }{\bf #1},}
\def\EPJC#1,{{\jnfont Eur.\ Phys.\ Jour.\ C }{\bf #1},}
\def\PRD#1,{{\jnfont Phys.\ Rev.\ D }{\bf #1},}
\def\PRL#1,{{\jnfont Phys.\ Rev.\ Lett.\ }{\bf #1},}
\def\MPLA#1,{{\jnfont Mod.\ Phys.\ Lett.\ A }{\bf #1},}
\def\JPG#1,{{\jnfont J.\ Phys.\ G }{\bf #1},}
\def\CTP#1,{{\jnfont Commun.\ Theor.\ Phys.\ }{\bf #1},}
\def\JHEP#1,{{\jnfont JHEP \ }{\bf #1},}
\def\NPPS#1,{{\jnfont Nucl.\ Phys.\ Proc.\ Suppl.\ }{\bf #1},}
\def\CPC#1,{{\jnfont Computl.\ Phys.\ Commun.\ }{\bf #1},}
\def\CPL#1,{{\jnfont Chin.\ Phys.\ Lett. }{\bf #1},}
\def\AJS#1,{{\jnfont Astrophys.\ J.\ Suppl. }{\bf #1},}
\def\PR#1,{{\jnfont Phys.\ Rept. }{\bf #1},}
\def\AP#1,{{\jnfont Astropart.\ Phys. }{\bf #1},}
\def\EPL#1,{{\jnfont Europhys.\ Lett. }{\bf #1},}
\def\FP#1,{{\jnfont Fortsch.\ Phys. }{\bf #1},}

\begin{document}

\title{\ \\[10mm] The recent Higgs boson data and Higgs triplet model with vectorlike quarks}

\author{Lei Wang, Xiao-Fang Han$^{ *}$\footnotetext{*) Corresponding author.
Email address: xfhan@itp.ac.cn}}

\affiliation{ Department of Physics, Yantai University, Yantai
264005, China}

%---------------------------------------------------------------------------

\begin{abstract}
Some vectorlike quarks are added to the Higgs triplet model with the
motivation of fitting the recent Higgs boson data released by LHC
and Tevatron collaborations. These vector-like quarks can suppress
the cross section of $gg\to h$ sizably, while the charged scalars,
especially for the doubly charged scalar, can enhance $Br(h\to
\gamma\gamma)$ more sizably. Besides, the Higgs couplings to $WW$,
$ZZ$ and light fermions can be the same as their SM values. Thus,
the model will enhance the Higgs production rates into
$\gamma\gamma$ and $jj\gamma\gamma$, while those for $WW^*$, $ZZ^*$
and $\tau\bar{\tau}$ at the LHC are reduced relative to their SM
predictions. The Higgs production rate into $Vb\bar{b}$ at the
Tevatron is the same as the SM prediction.

\end{abstract}



\maketitle

\section{Introduction}

The hint of a Higgs particle around 125 GeV revealed earlier by the
ATLAS and CMS experiments \cite{lhc,atlas,cms,jjh}, has now become
an indisputable discovery \cite{7.4}, which is supported by the data
from the Tevatron collider experiments \cite{tevatron}. Although the
present results of all the search channels have large uncertainty,
all measured $\gamma\gamma$ rates have central values above the
standard model (SM) prediction, and all the $WW^*$ rates have
central values below the SM prediction
\cite{7.4,cmsww,moriondcms,moriondatlas}. The observation of the
channel $q\bar{q}\to Vh$ with $h\to b\bar{b}$ at the Tevatron
disfavors the scenarios in which the Higgs couplings to gauge bosons
and bottom quarks are significantly reduced with respect to their SM
values \cite{12041252}. Ref. \cite{12034254} performed a
phenomenological fit to the new ATLAS, CMS, CDF and D0 Higgs boson
data, and found that the present data favor a reduction of $gg\to h$
rate and an enhancement of $h\to \gamma\gamma$ rate.

In this paper, we introduce some vector-like quarks in the framework
of Higgs triplet model (HTM) with the motivation of obtaining a
Higgs boson with the properties mentioned above. The HTM which we
will study contains a complex doublet Higgs field and a complex
triplet Higgs field with hypercharge Y = 2 \cite{htm}. Several
physical Higgs bosons remain after the spontaneous symmetry
breaking, including two CP-even ($h$ and $H$), one CP-odd ($A$), one
charged ($H^\pm$) and one doubly charged Higgs scalars
($H^{\pm\pm}$). We will take $h$ as a purely SM-like Higgs with
around 125 GeV mass, and its couplings to $WW$, $ZZ$ and light
fermions nearly equal to their SM values, which can fit the
$Vb\bar{b}$ rate measured at the Tevatron relatively well. The rate
of $gg\to h$ is sizably suppressed by the vector-like quarks while
the $Br(h\to\gamma\gamma)$ is more sizably enhanced by the singly
and doubly charged scalars. The LHC Higgs diphoton signal has been
studied in the original HTM \cite{htmrr,htmrr2}. The recent Higgs
data has been discussed in various extensions of the SM, such as the
minimal supersymmetric standard model (MSSM) \cite{11125453-26272829},
 the next-to-MSSM \cite{11125453-3537},
 the inert Higgs doublet model \cite{12012644}, the two Higgs doublet model
\cite{11125453-3234,12071083}, the little Higgs models
\cite{11125453-3839}, the models with extra dimension \cite{extrad}
and the models with the universal varying Yukawa couplings
\cite{12035083}.

 This work is organized as follows. In Sec. II, we briefly
review the Higgs triplet model and then introduce some vector-like
quarks. In Sec. III, we discuss the $gg\to h$ rate and branching
ratio of $h\to \gamma\gamma$. In Sec. IV, we calculate the Higgs
production rates at the LHC and Tevatron. Finally, we give our
conclusion in Sec. V.

\section{The Higgs triplet model with vector-like quark}
In the HTM, a complex $\rm{SU(2)_L}$ triplet scalar field $\Delta$ with Y = 2
is added to the SM Lagrangian in addition to the doublet field $\Phi$.
These fields can be written as
\begin{eqnarray}
\Delta &=\left(
\begin{array}{cc}
\delta^+/\sqrt{2} & \delta^{++} \\
\delta^0 & -\delta^+/\sqrt{2}\\
\end{array}
\right),  \qquad \Phi=\left(
                    \begin{array}{c}
                      \phi^+ \\
                      \phi^0 \\
                    \end{array}
                  \right).
\end{eqnarray}
 The renormalizable scalar potential can
be written as
\begin{eqnarray}\label{potent}
V&=&-m_\Phi^2{\Phi^\dagger{\Phi}}+\frac{\lambda}{4}(\Phi^\dagger{\Phi})^2+
M_\Delta^2Tr(\Delta^{\dagger}{\Delta}) +
\lambda_1(\Phi^\dagger{\Phi})Tr(\Delta^{\dagger}{\Delta}) \\ & + &
\lambda_2(Tr\Delta^{\dagger}{\Delta})^2
+\lambda_3Tr(\Delta^{\dagger}{\Delta})^2 +
\lambda_4{\Phi^\dagger\Delta\Delta^{\dagger}\Phi}+
[\mu(\Phi^T{i}\tau_2\Delta^{\dagger}\Phi)+h.c.].\nonumber
\end{eqnarray}
The Higgs doublet and triplet field can acquire vacuum
expectation values
\begin{equation}
\langle \Phi \rangle = \frac{1}{\sqrt{2}} \left(
                    \begin{array}{c}
                      0 \\
                      v_d \\
                    \end{array}
                  \right), \qquad \langle \Delta \rangle = \frac{1}{\sqrt{2}}
\left(
\begin{array}{cc}
0 & 0 \\
v_t & 0\\
\end{array}
\right)
\label{vacuum}
\end{equation} with $v^2=v_d^2+4v_t^2\approx(246~\rm{GeV})^2$.

In the HTM, there are seven physical Higgs bosons, including two
CP-even ($h$ and $H$), one CP-odd ($A$), one charged ($H^\pm$) and
one doubly charged Higgs scalars ($H^{\pm\pm}$). The scalar
potential contains seven independent parameters: $\lambda$,
$\lambda_{i=1...4}$, $\mu$  and $v_t$. Since the experimental value of
the $\rho$ parameter is near unity, $v^2_d/v^2_t$ is required to be
much smaller than unity at tree-level, which can produce naturally a
very small neutrino mass for Yukwawa coupling of order 1 \cite{ro}.
The perturbative unitarity and potential boundedness can give strong
constraints on these parameters.  Refer to \cite{htmrr}, we take
\bea {\it v}_t&=&\mu=1\,{\rm GeV},\qquad \lambda = 0.516,
\qquad 0 < \lambda_1 < 10, \nonumber\\
\lambda_3&=&2\lambda_2 = 0.2,
\qquad -2 < \lambda_4 < 1.
\label{space}\eea
For such parameter space, the seven Higgs masses can be given as
\begin{eqnarray}
m_{h}^2&\simeq&\frac{\lambda}{2}v^2_d\simeq (125~\rm{GeV})^2, \nonumber\\
m_{H}^2&\simeq&m_{A}^2\simeq\frac{\sqrt{2}\mu v_d^2}{2v_t}, \nonumber\\
m_{H^{\pm\pm}}^2&=&\frac{\sqrt{2}\mu{v_d^2}- \lambda_4v_d^2v_t-2\lambda_3v_t^3}{2v_t},
\nonumber \\
m_{H^\pm}^2&=&\frac{(v_d^2+2 v_t^2)[2\sqrt{2}\mu- \lambda_4 v_t]}{4v_t}\simeq m_{H^{\pm\pm}}^2+\frac{\lambda_4}{4}v_d^2.  \label{mass}
\end{eqnarray}
Where we take $\mu={\it v}_t$ in order to make $h$ to be a purely
SM-like Higgs boson, for which the mixing of $h$ and $H$ is nearly
absent. The cosine value of the mixing angle is always larger than
0.996 for the parameter space shown in Eq. (\ref{space}). $\lambda =
0.516$ determines the mass of $h$ to be around 125 GeV. When $\mu$
is much less than $v_t$, $h$ and $H$ have large mixing. The $H$ can
be even as a purely SM-like Higgs boson for enough small $\mu$ \cite{htmrr}.
However, the large $\mu$ can enhance the masses of $H^+$ and
$H^{++}$, which will suppress their contributions to $h\to
\gamma\gamma$ sizably.

The scalar potential term in Eq. (\ref{potent}) contains the SM-like
Higgs boson coupling to the charged scalars \cite{htmrr}, \beq
g_{h H^{++}H^{--}} \approx   -  \lambda_1v_d,\qquad g_{h H^+H^-}
\approx - (\lambda_{1} + \frac{\lambda_{4}}{2}) v_d.
\label{eq:gcalHHp} \eeq

Now we add some vector-like quarks to the HTM. When a unique
additional vector-like multiplet are introduced to the SM
Lagrangian, the cross section of $gg\to h$ will be increased or
slightly decreased relatively to the SM prediction since the
physical signs of the Yukawa couplings are identical for the top
quark and extra quark. Ref. \cite{vector} proposed a minimal
scenario in which the cross section of $gg\to h$ can be strongly
suppressed. Following the approach in ref. \cite{vector}, we
introduce the doublet $\left(q_{5/3},t' \right)^T_{L,R}$, the
singlet $t''_L$  and $t''_R$ in addition to the SM-like fields.
Where the $L/R$ represents the fermion chirality and $q_{5/3}$ is an
exotic quark with the electric charge ${5/3}$. Their Yukawa
interactions with the doublet Higgs field can be written as
\begin{equation}
{\cal L}_{\rm Yuk} =
y \overline{\left ( \begin{array}{c}
t \\  b
\end{array} \right )}_{L}
\tilde{\Phi}~t_R +
y' \overline{\left ( \begin{array}{c}
q_{5/3} \\  t'
\end{array} \right )}_L
\Phi~t_R +
y'' \overline{\left ( \begin{array}{c}
q_{5/3} \\  t'
\end{array} \right )}_{L/R}
\Phi~t''_{R/L} +
\tilde y \overline{\left ( \begin{array}{c}
t \\ b
\end{array} \right )}_L
\tilde{\Phi}~t''_R
\nonumber
\end{equation}
\begin{equation}
+ \ y_b \overline{\left ( \begin{array}{c}
t \\  b
\end{array} \right )}_{L}
\Phi~b_R
+ m \ \bar t''_L~t_R
+ m' \overline{\left ( \begin{array}{c}
q_{5/3} \\  t'
\end{array} \right )}_{L}
\left ( \begin{array}{c}
q_{5/3} \\  t'
\end{array} \right )_{R}
+ m'' \ \bar t''_{L}~t''_{R}
+ {\rm h.c.,}
\label{VTH:LagDoub}
\end{equation}
where $\tilde{\Phi}=i\sigma_2 \Phi^*$. After EWSB takes place, the Lagrangian~(\ref{VTH:LagDoub}) gives rise  to
the top mass matrix:
\begin{equation}
{\cal L}_{\rm mass} =
\overline{\left ( \begin{array}{c} t \\ t' \\ t'' \end{array}
\right )}_L
\left ( \begin{array}{ccc}
yv/\sqrt{2} & 0 & \tilde y v/\sqrt{2}\\
y' v/\sqrt{2} & m' & y'' v/\sqrt{2}\\
m & y'' v/\sqrt{2} & m''
\end{array} \right )
\left ( \begin{array}{c} t^c \\ t' \\ t'' \end{array} \right )_R+
h.c. \label{VTH:MassDoub}
\end{equation}
The Yukawa coupling $y''$ sign of top partner $t'$ and $t''$ can be
taken independently of the top quark Yukawa coupling $y$ sign in
order to generate destructive interferences between the top quark
and top parter contributions to $gg\to h$. After diagonalizing the
mass matrix, we can get the mass eigenstates $t$, $t_1$ and $t_2$ as
well as their couplings with the Higgs boson. The exotic quark
$q_{5/3}$ has no the coupling to $h$ at tree-level.

The triplet Higgs field $\Delta$ can mediate the interactions
between the right-handed doublet quark field with Y = 7/3 and the
left-handed doublet quark field with Y = 1/3. These interactions
affect hardly the $h$ production rates since the mixing angle between
$h$ and $H$ is taken as very small in order to make $h$ to be a
purely SM-like Higgs. In the same way, the $h$ couplings to the
gauge bosons and light fermions equal to their SM values nearly.

\section{$gg\to h$ and $h\to \gamma\gamma$}
In the SM, the main production processes of Higgs boson at the LHC
include gluon-gluon fusion ($gg \to h$), vector-boson fusion (VBF)
and associated production with $W$ and $Z$ bosons (Vh). Their cross
sections are respectively \cite{cs}\bea
&& \sigma(gg\to h)= (15.3 \pm 2.6)~\rm{pb},~~~~\sigma(pp\to jjh)= 1.2~\rm{pb} \nonumber\\
&& \sigma(pp\to Wh)= 0.57~\rm{pb},~~~~\sigma(pp\to Zh)= 0.32
~\rm{pb}.\eea Compared with SM, as a purely SM-like Higgs with 125 GeV
mass, only $h\leftrightarrow gg$ and $h\to \gamma\gamma$ at one-loop
 are modified in the HTM with vector-like quark (HTMVQ), and
the rates for other processes at tree-level are the same as the SM
predictions.

\subsection{The cross section of $gg\to h$}
At the LHC the cross section of the single Higgs production via
gluon-gluon fusion can be given,
\bea
&&
\sigma(gg\to h)\nonumber =\tau_0\int_{\tau_0}^1
\frac{dx}{x}f_{g}(x,~\mu^2_F)f_{g}(\frac{\tau_0}{x},~\mu^2_F)\hat{\sigma}(gg\to
h) ~\rm{with} \nonumber\\
&&\hat{\sigma}(gg\to h)=\Gamma(h\to gg)\frac{\pi^2}{8m^3_h},
\label{gghhgg}
\eea
where $\tau_0=\frac{m_h^2}{s}$ with $\sqrt{s}$
being the center-of-mass energy of the LHC and $f_{g}(x,~\mu^2_F)$
is the parton distributions of gluon. The Eq. (\ref{gghhgg}) shows
that the $\sigma(gg\to h)$ has a strong correlation with the decay
width $\Gamma(h\to gg)$.

In the SM, $\Gamma(h\to gg)$ is dominated by top quark loop. The
HTMVQ gives the corrections via the modified couplings $ht\bar{t}$
and  the loops of top partner $t_1$ and $t_2$. In the HTMVQ,
$\Gamma(h\to gg)$ can be written as \cite{hrr1loop}
\begin{eqnarray}
\Gamma(h\to gg) = \frac{\alpha_s^2
  m^3_{h}}{128\pi^3v^2}\Bigg| y_t F_{1/2}(\tau_t)
+ y_{t_1} F_{1/2}(\tau_{t_1})+ y_{t_2} F_{1/2}(\tau_{t_2}) \Bigg|^2 ,
\label{widgg}
\end{eqnarray}
where $\tau_{f}=\frac{4m_f^2}{m_h^2}$. The expression of $F_{1/2}$
is given in Eq. (\ref{ffun}).

%%%%%%%%%%%%%%%%%%%%%%%%%%
\begin{table}[!ht]
\vspace*{.5cm}
\begin{center}
\begin{tabular}{|c|c|c|c|c|c|c|c|c|c|c|c|c|c|}
\hline
Parameter Set &A1 &A2& B1 & B2 &  C1  & C2 &D1 &D2 &E1 & E2\\
\hline
$y $   & 1.215 & 1.226 & 1.144   & 1.200  & 1.164 & 1.092 & 1.087 & 1.107 & 1.045 & 1.039\\
\hline
$y'$ & -0.866 & -1.124 & -0.842  & -1.386 & -1.219  & -0.753 & -0.855 & -1.009 &-0.749&-0.983\\
\hline
$\tilde{y}/y$ &0.705 & 0.898& 0.798  & 0.818  & 0.380  & 0.791 & 0.540 & 0.890 &0.568&0.568\\
\hline
$y''/y'$ &-1.870& -1.546& -1.922  & -1.169  & -1.069  & -1.944 & -1.341 & -1.148 &-1.008&-1.026\\
\hline
$m'$ (GeV) &507.3 & 669.7& 609.5  & 735.0  &  552.3 & 715.7 & 554.0 & 711.8 &547.8&960.0\\
\hline
\hline
$m_{t_1}$ (GeV) & 428.9&546.9&498.7 & 615.4 & 510.8 & 589.1 & 506.5 & 641.2 &532.7&819.3\\
\hline
$m_{t_2}$ (GeV) & 1136 & 1193 &1156 & 1193 & 1099 & 1158 & 1081.0 & 1115.3 &1040&1164\\
\hline
$y_{t}$ &0.362&0.380&0.567&0.466& 0.604&0.765 & 0.754 & 0.728 &0.852&0.899\\
\hline
$y_{t_1}$ &-0.110&-0.171&-0.231& -0.151& -0.050&-0.261& -0.109 & -0.118 &-0.028&-0.165\\
\hline
$y_{t_2}$ &0.199&0.241&0.213&0.234& 0.157&0.204& 0.131 & 0.165 &0.072&0.161\\
\hline
 $R_{gg}$ &0.201&0.200& 0.302 & 0.299 & 0.501 & 0.502& 0.601 & 0.598 &0.802& 0.801\\
\hline $\sigma_{\bar t_2t_2 \to b W\bar{b} W}$ (pb) &0.0232
&0.0041&0.0073   & 0.0016 &
0.0040 & 0.0025  &  0.0048   & 0.0013  &0.0028&0.0002\\
LHC bound~\cite{bw} &$<$ 0.22 & $<$ 0.16 &$<$ 0.19 & $\times$ & $<$ 0.18  & $<$ 0.14 & $<$ 0.14 & $\times$ & $<$ 0.17& $\times$\\
\hline $\sigma_{\bar t_2t_2 \to t Z\bar{t}Z}$ (pb) &0.0321&0.0060&
0.0118 & 0.0030 &
0.0153 & 0.0041 & 0.0173 & 0.0030 &0.0166&0.0004\\
LHC bound~\cite{tz} &$<0.29$ & $<0.28$ &$<0.30$ & $\times$ & $< 0.30 $  & $\times$ & $<$ 0.3 & $\times$ & $<$ 0.29 & $\times$ \\
\hline
\end{tabular}
\end{center}
\caption{The several points for $R_{gg}\simeq$ 0.2, 0.3, 0.5, 0.6,
and 0.8. In addition to the parameters shown above, we take $m=0$
GeV, $m''=1000$ GeV, $m_h =125$ GeV and require $m_t\simeq172.5$
GeV. $y_f=\frac{v}{m_{f}}g_{hf\bar{f}}$ with $g_{hf\bar{f}}$ being
the coupling constant of $hf\bar{f}$. $\sigma_{\bar t_2t_2 \to b
W\bar{b} W}$ and $\sigma_{\bar t_2t_2 \to t Z\bar{t}Z}$ represent
$\sigma(pp\to t_1 \bar{t}_1)\times Br^2(t_1\to bW)$ and
$\sigma(pp\to t_1 \bar{t}_1)\times Br^2(t_1\to tZ)$ , respectively.}
\label{tabgg}
\end{table}

In Table \ref{tabgg}, we list several points for $R_{gg} \equiv
\frac{\sigma(gg\to h)}{\sigma(gg\to h)_{SM}}$ $\simeq$ 0.2, 0.3,
0.5, 0.6 and 0.8, respectively. The top partner $t_1$ mainly decays
into $th$, $tZ$ and $bW$. The CMS experiments at the LHC have
released the results of their searches for vector-like quark, and
give the upper bounds of $\sigma(pp\to t_1 \bar{t}_1)\times
Br^2(t_1\to bW)$ and $\sigma(pp\to t_1 \bar{t}_1)\times Br^2(t_1\to
tZ)$. The HTMVQ predictions and LHC upper bounds for these rates are
given in Table \ref{tabgg}. $\sigma(pp\to t_1 \bar{t}_1)$ is
calculated with the HATHOR program \cite{HATHOR} at NNLO. The cross
section of $t_2\bar{t}_2$ at the LHC can be severely suppressed by
the large mass (over 1 TeV), which can be free from the constraints
of LHC direct searches experiments. From Table \ref{tabgg}, we see
that, being in agreement with the experimental constraints of LHC
direct searches for $t_1$ and $t_2$, the two vector-like top
partners can suppress the cross section of $gg\to h$ sizably, and
the cross section can be reduced by a factor of 0.2. The reduced top
Yukawa coupling and  the opposite sign between the Yukawa couplings
of top and $t_1$ are responsible for the suppression of
$\sigma(gg\to h)$.

\subsection{the branching ratio of $h\to \gamma\gamma$}

In the SM, the decay $h\to \gamma\gamma$ is dominated by the $W$ loop
which can interfere destructively with the subdominant top
quark loop. In the HTMVQ, the singly charged scalar $H^{\pm}$, the
doubly charged scalar $H^{\pm\pm}$, top partner $t_1$ and $t_2$ will
give the additional contributions to the decay width $\Gamma(h\to
\gamma\gamma)$, which can be expressed as \cite{hrr1loop}
\begin{eqnarray}
\Gamma(h\to \gamma\gamma) = \frac{\alpha^2
  m^3_{h}}{256\pi^3v^2}\Bigg|  F_{1}(\tau_W)+ \sum_{i} N_{cf} Q^2_{f} y_{f} F_{1/2}(\tau_f)
+  g_{_{H^{\pm}}}F_{0}(\tau_{H^\pm})+  4g_{_{H^{\pm\pm}}}F_{0}(\tau_{H^{\pm\pm}})\Bigg|^2 ,
\label{gamrr}
\end{eqnarray}
where \bea &&\tau_W=\frac{4m_W^2}{m_h^2},~~~
\tau_{H^\pm}=\frac{4m_{H^\pm}^2}{m_h^2},~~~\tau_{H^{\pm\pm}}=\frac{4m_{H^{\pm\pm}}^2}{m_h^2},\nonumber\\
&&g_{_{H^{\pm}}}=-\frac{v}{2m_{H^\pm}^2}g_{hH^+ H^-},~~~
g_{_{H^{\pm\pm}}}=-\frac{v}{2m_{H^{\pm\pm}}^2}g_{hH^{++}
H^{--}}.\eea $N_{cf}$, $Q_f$ are the color factor and the electric
charge respectively for fermion $f$ running in the loop. The dimensionless loop factors for particles of spin
given in the subscript are:
\begin{eqnarray}
F_1 = 2+3\tau + 3\tau(2-\tau)f(\tau), \quad F_{1/2} =
-2\tau[1+(1-\tau)f(\tau)], \quad F_0 = \tau[1-\tau f(\tau)],
\label{ffun}\end{eqnarray}
with
\begin{equation}
f(\tau) = \left\{ \begin{array}{lr}
[\sin^{-1}(1/\sqrt{\tau})]^2, & \tau \geq 1 \\
-\frac{1}{4} [\ln(\eta_+/\eta_-) - i \pi]^2, & \, \tau < 1
\end{array}  \right.
\end{equation}
where $\eta_{\pm} = 1 \pm \sqrt{1-\tau}$.

\begin{figure}[tb]
%\begin{center}
 \epsfig{file=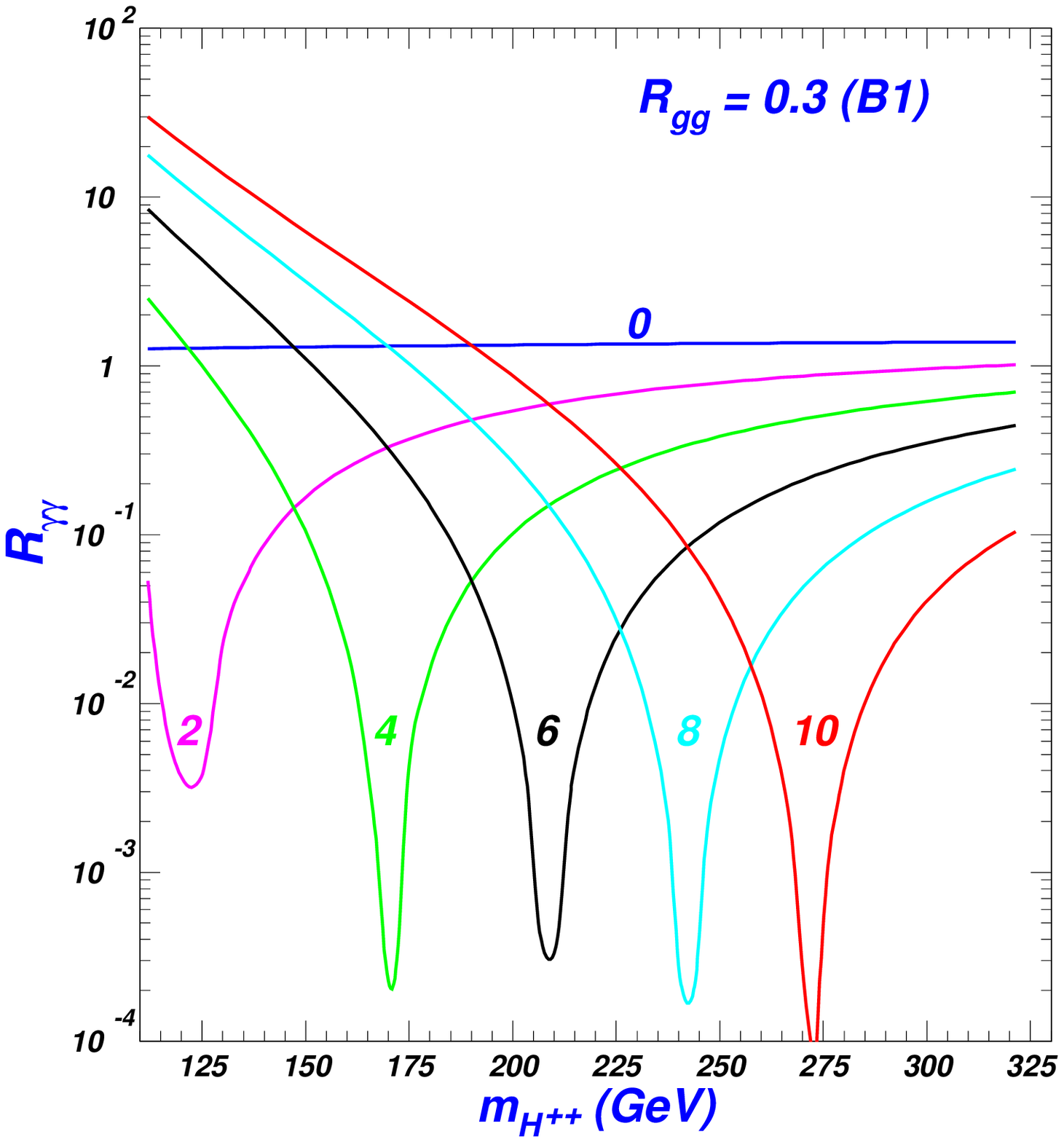,height=6.01cm}
  \epsfig{file=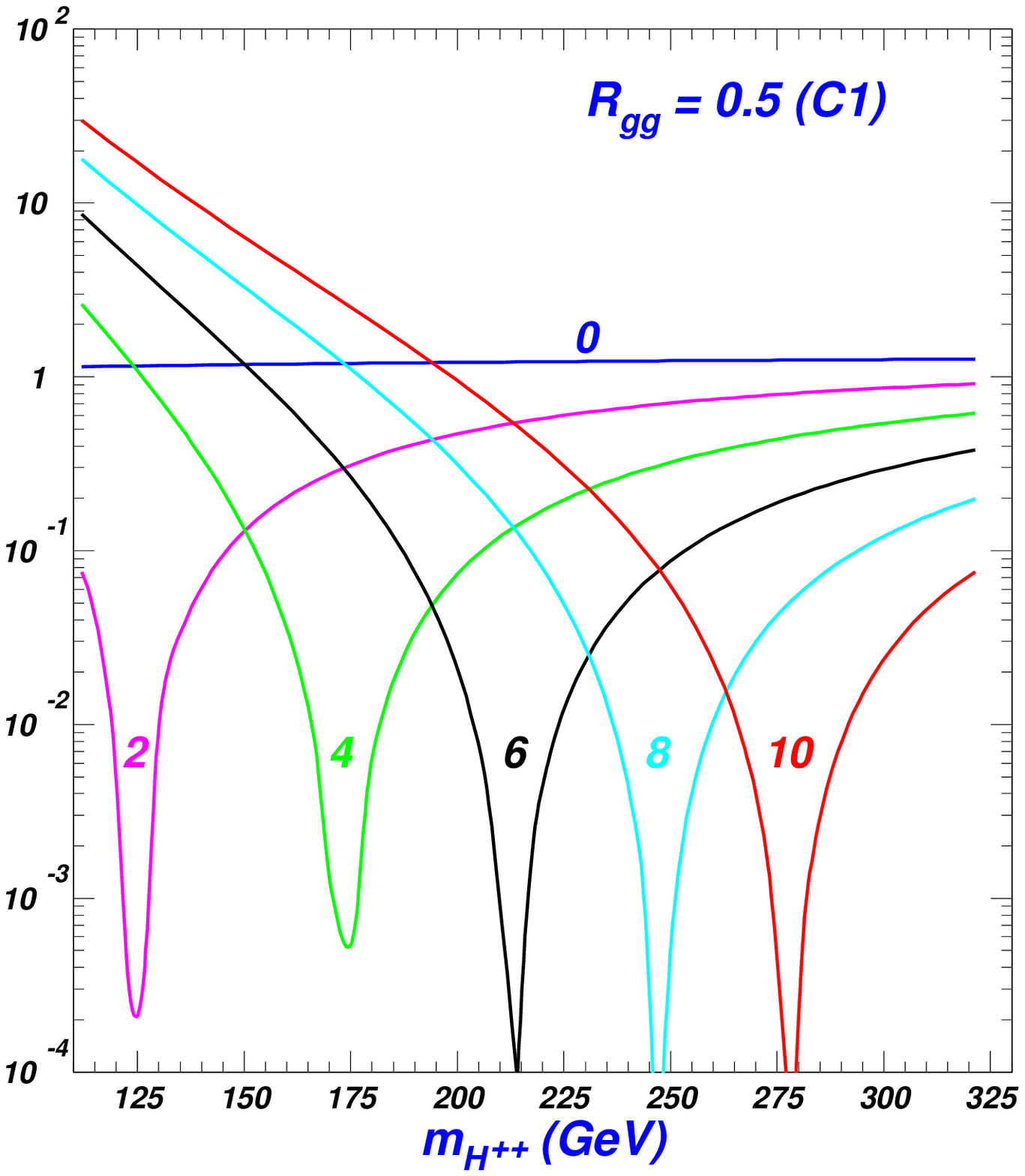,height=6.01cm}
  \epsfig{file=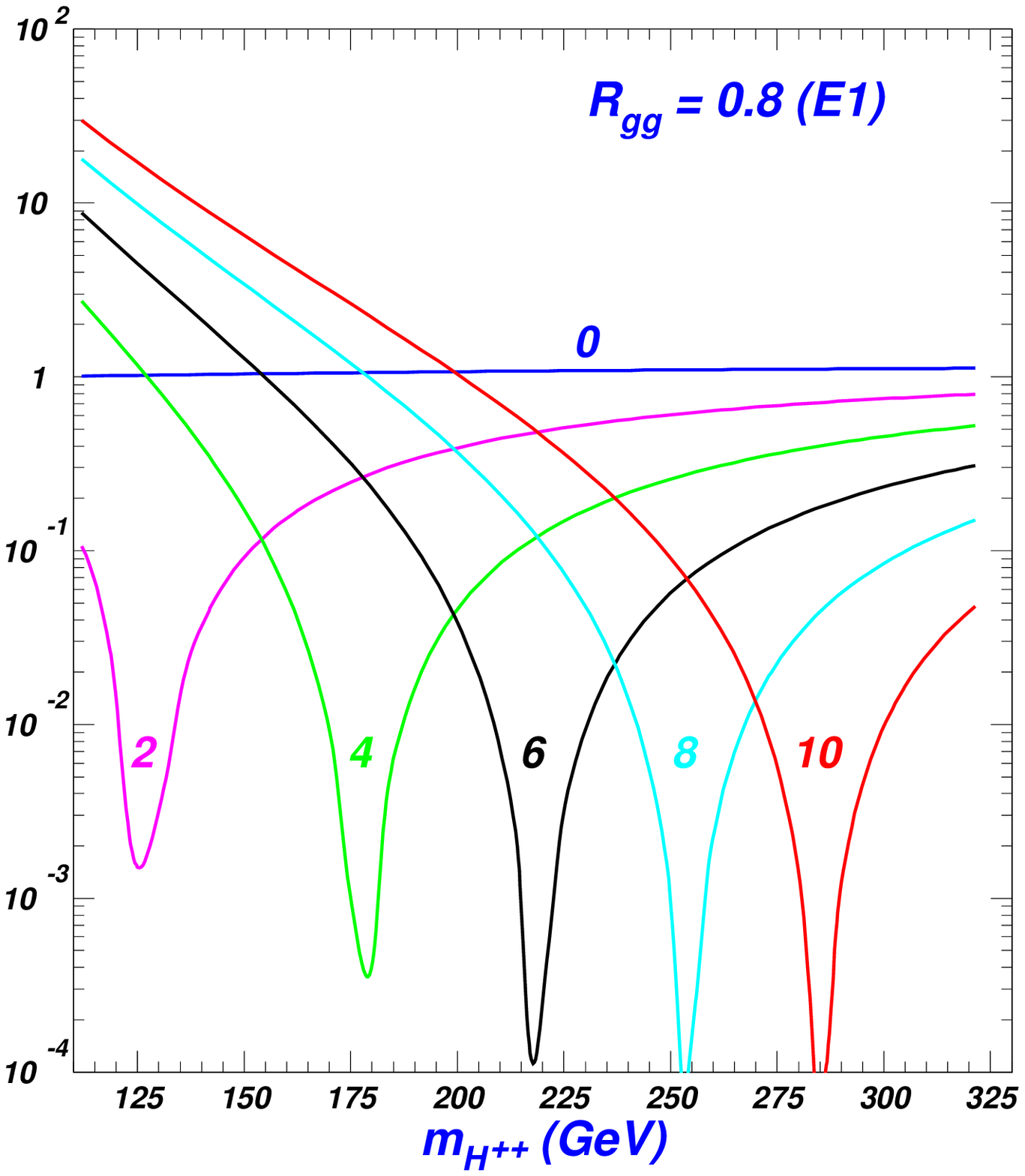,height=6.01cm}
    %\end{center}
\vspace{-0.4cm} \caption{$R_{\gamma\gamma}$ versus $m_{H^{\pm\pm}}$
for $R_{gg}=$ 0.3, 0.5 and 0.8, respectively. The numbers on the
curves denote the coupling constant $\lambda_1$.} \label{rrrmass}
\end{figure}

\begin{figure}[tb]
%\begin{center}
\epsfig{file=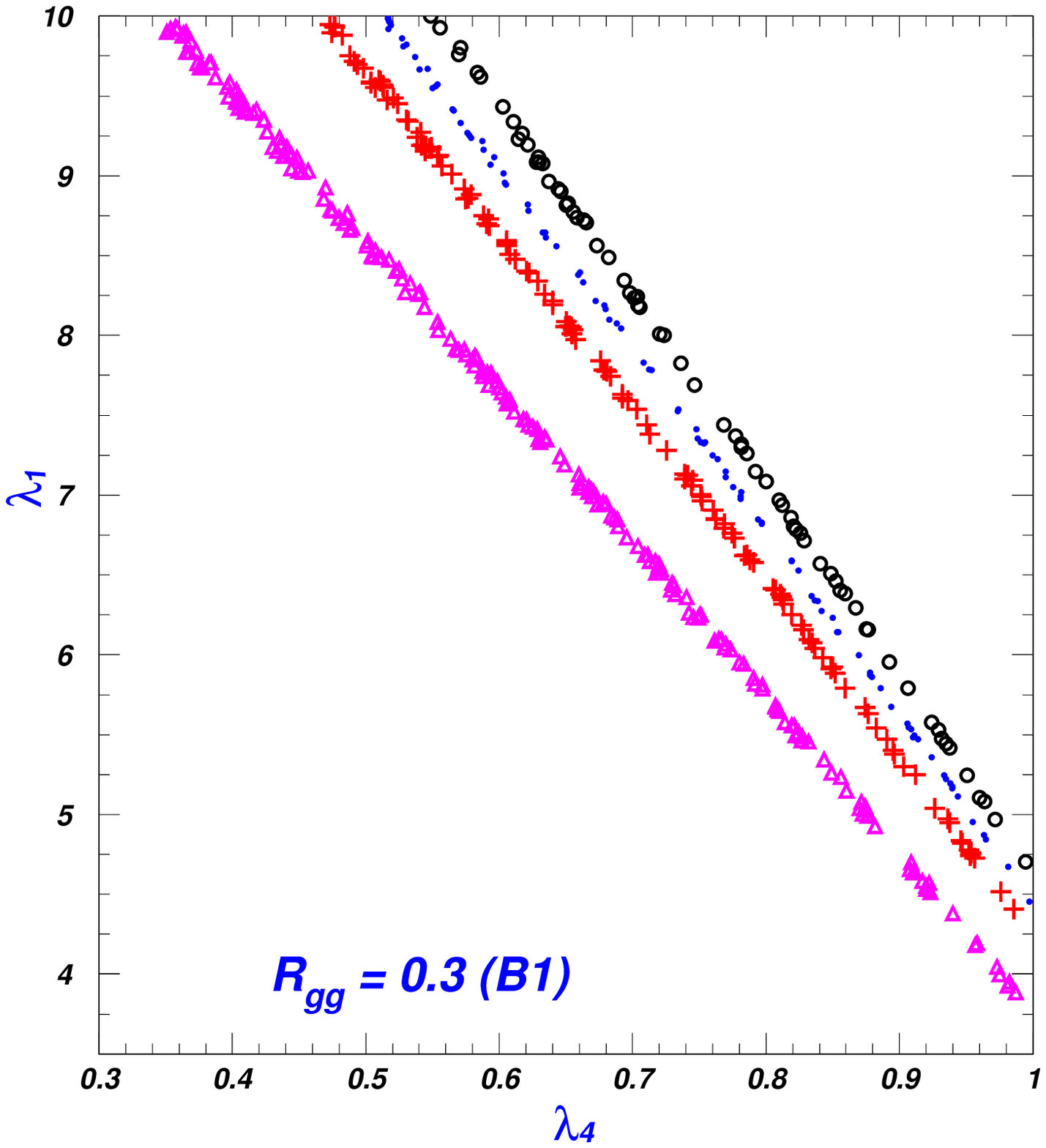,height=6.05cm}
    \epsfig{file=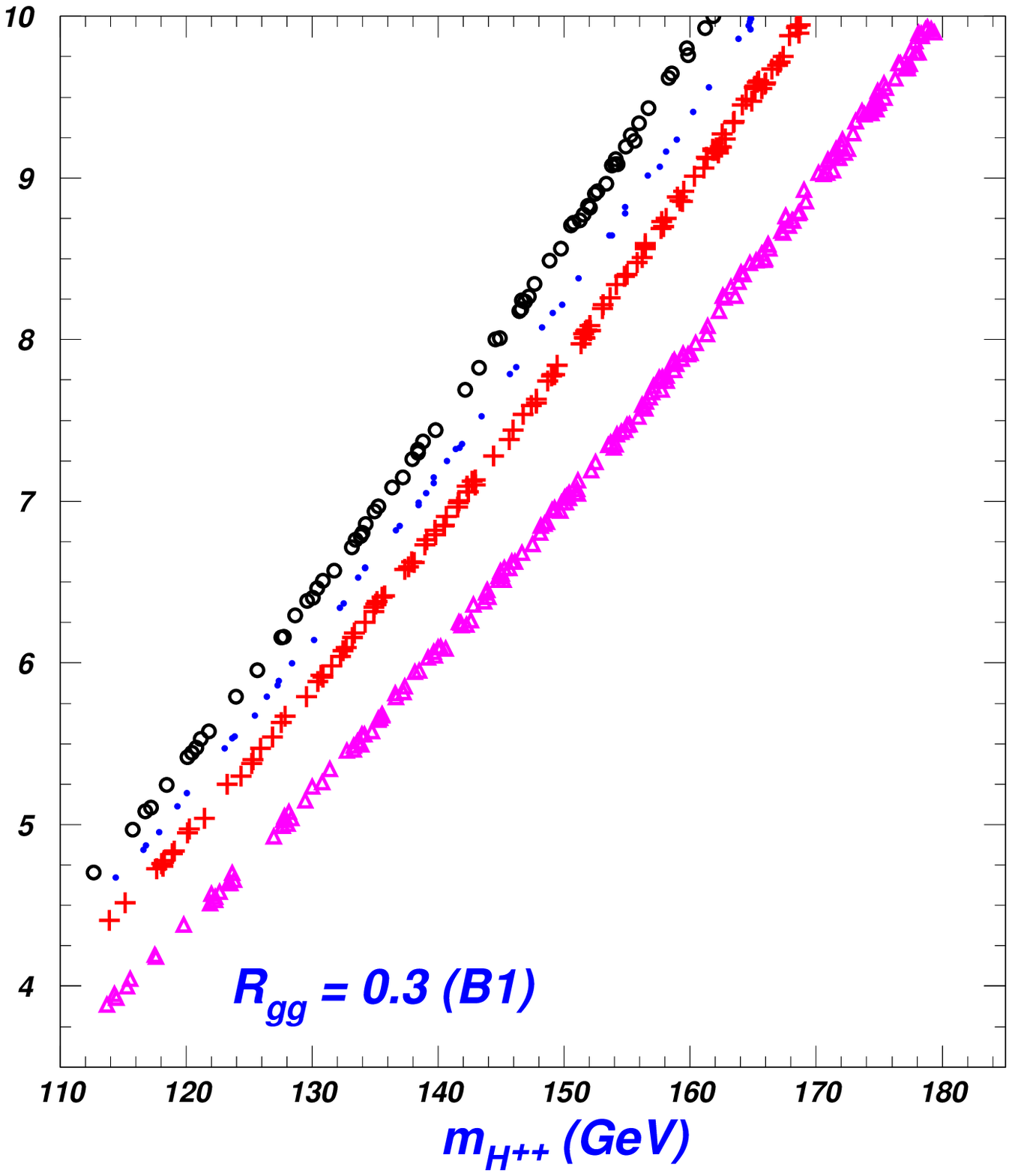,height=6.05cm}
    \epsfig{file=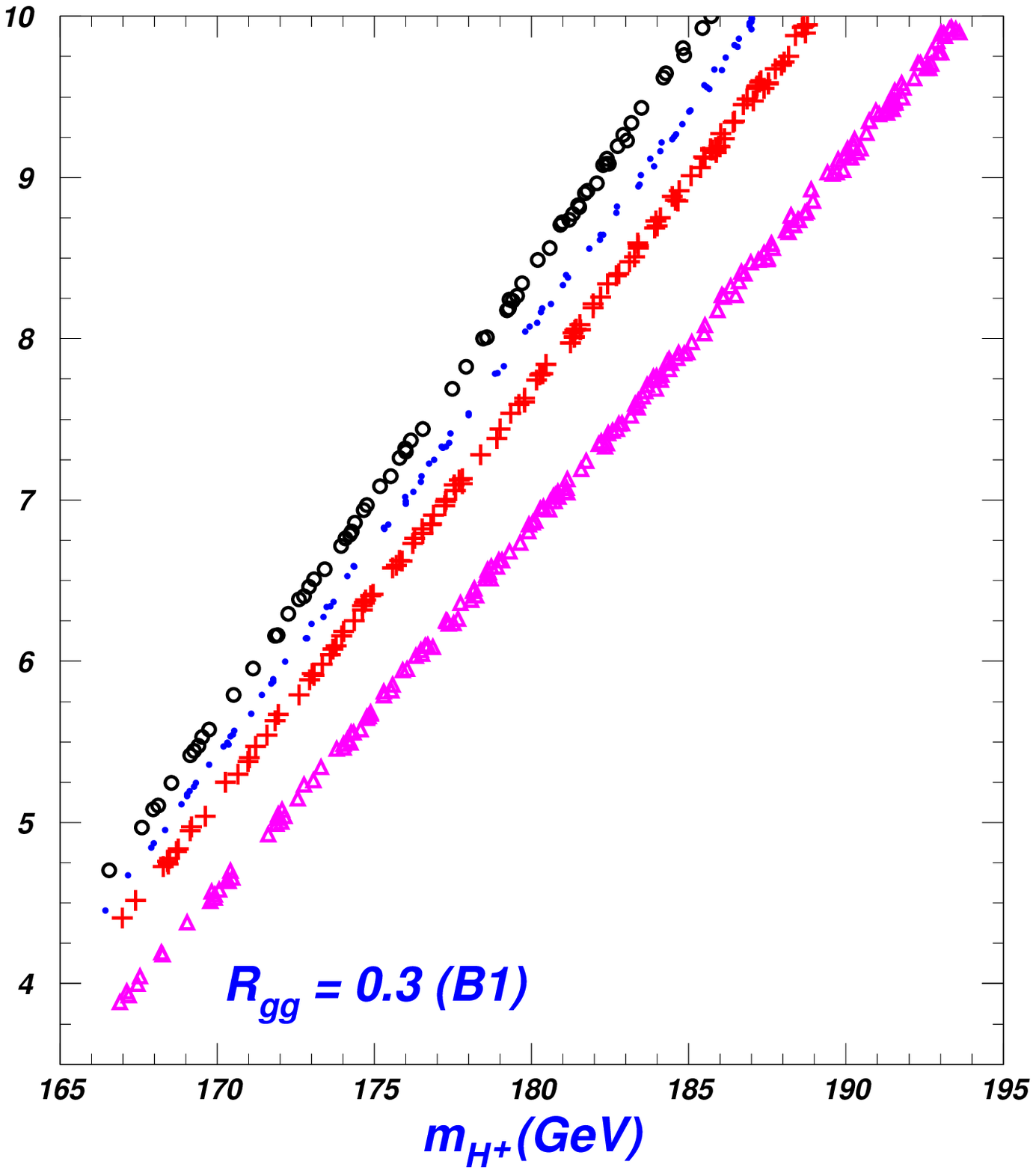,height=6.05cm}
    \epsfig{file=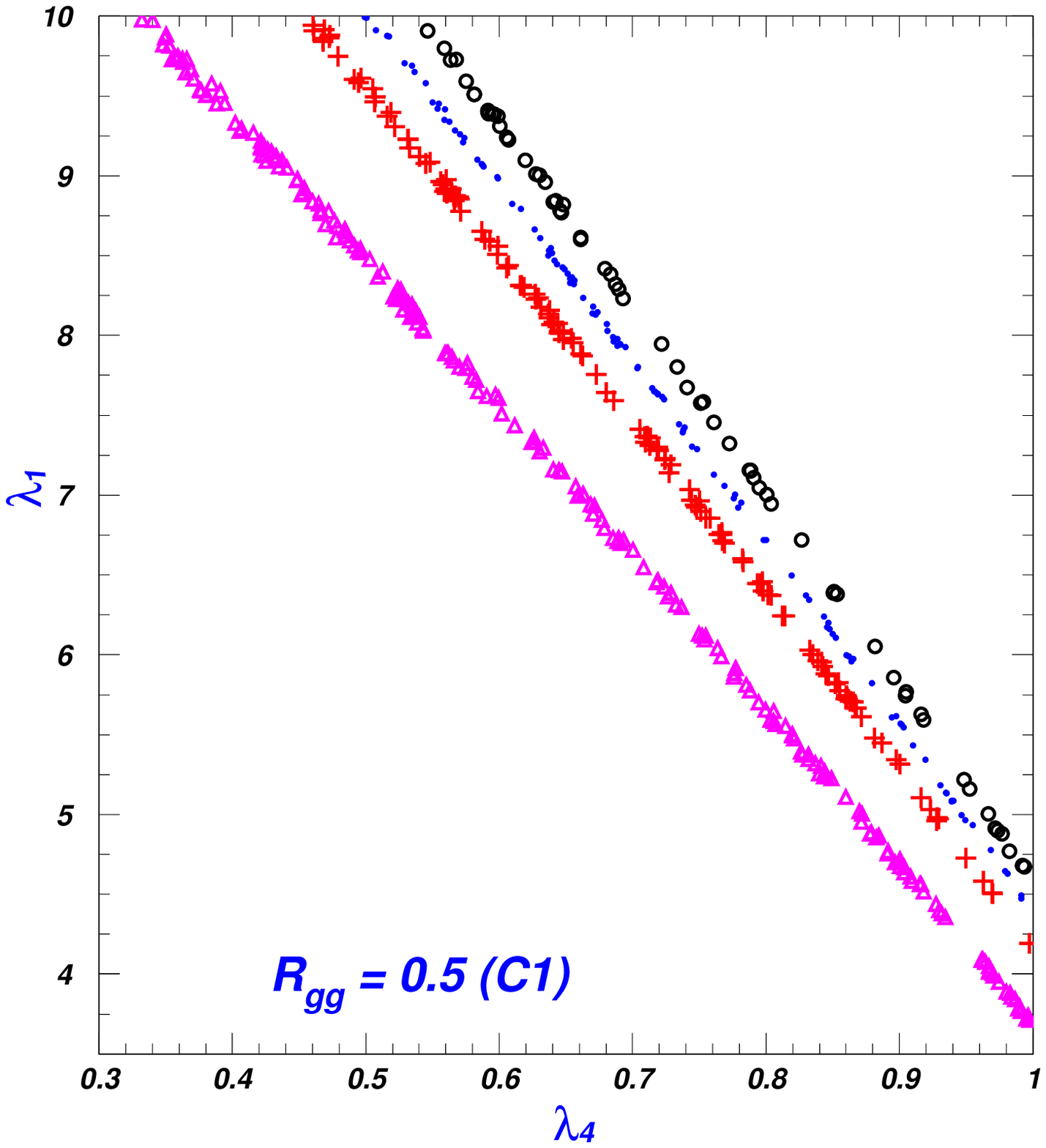,height=6.05cm}
     \epsfig{file=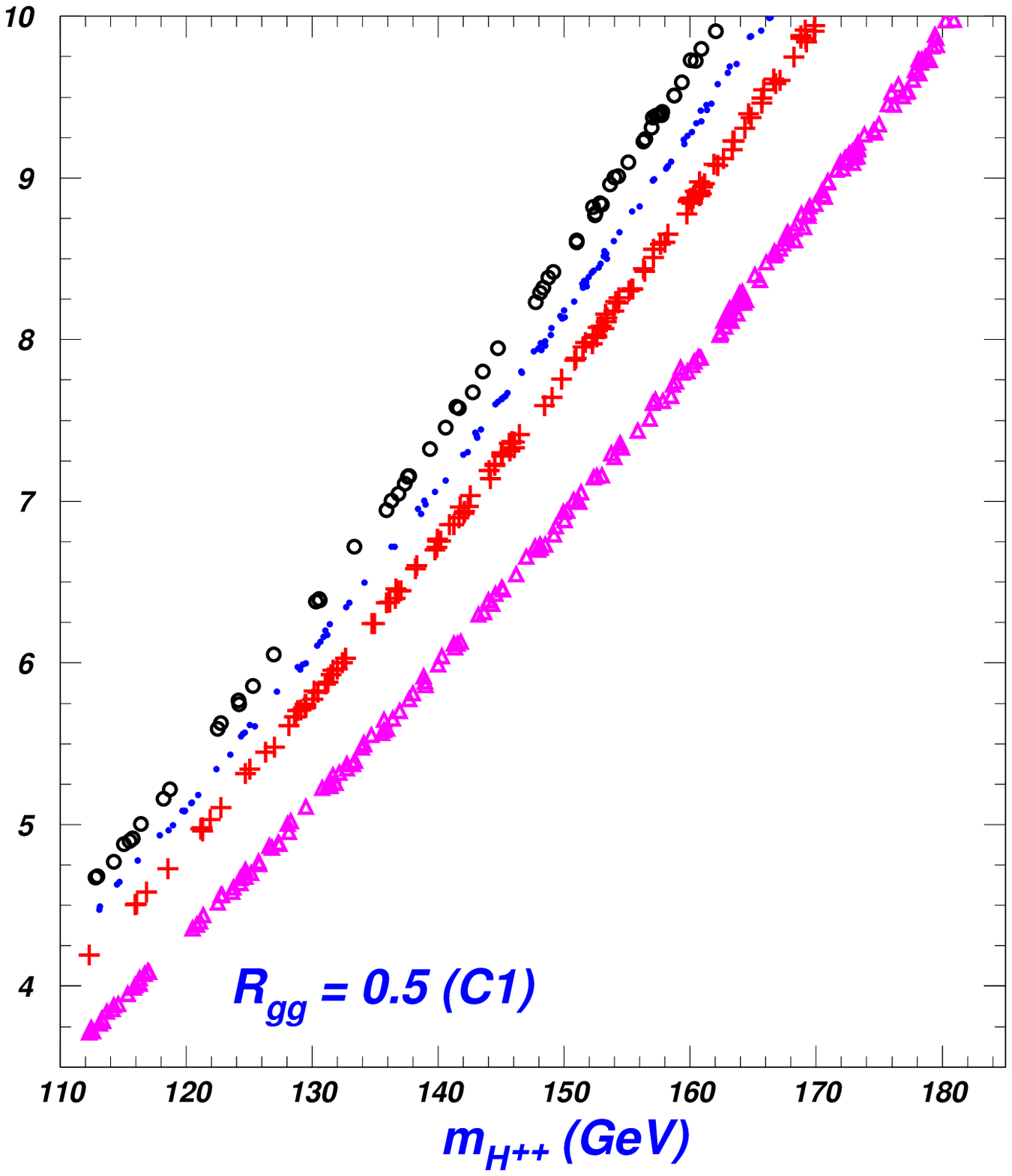,height=6.05cm}
  \epsfig{file=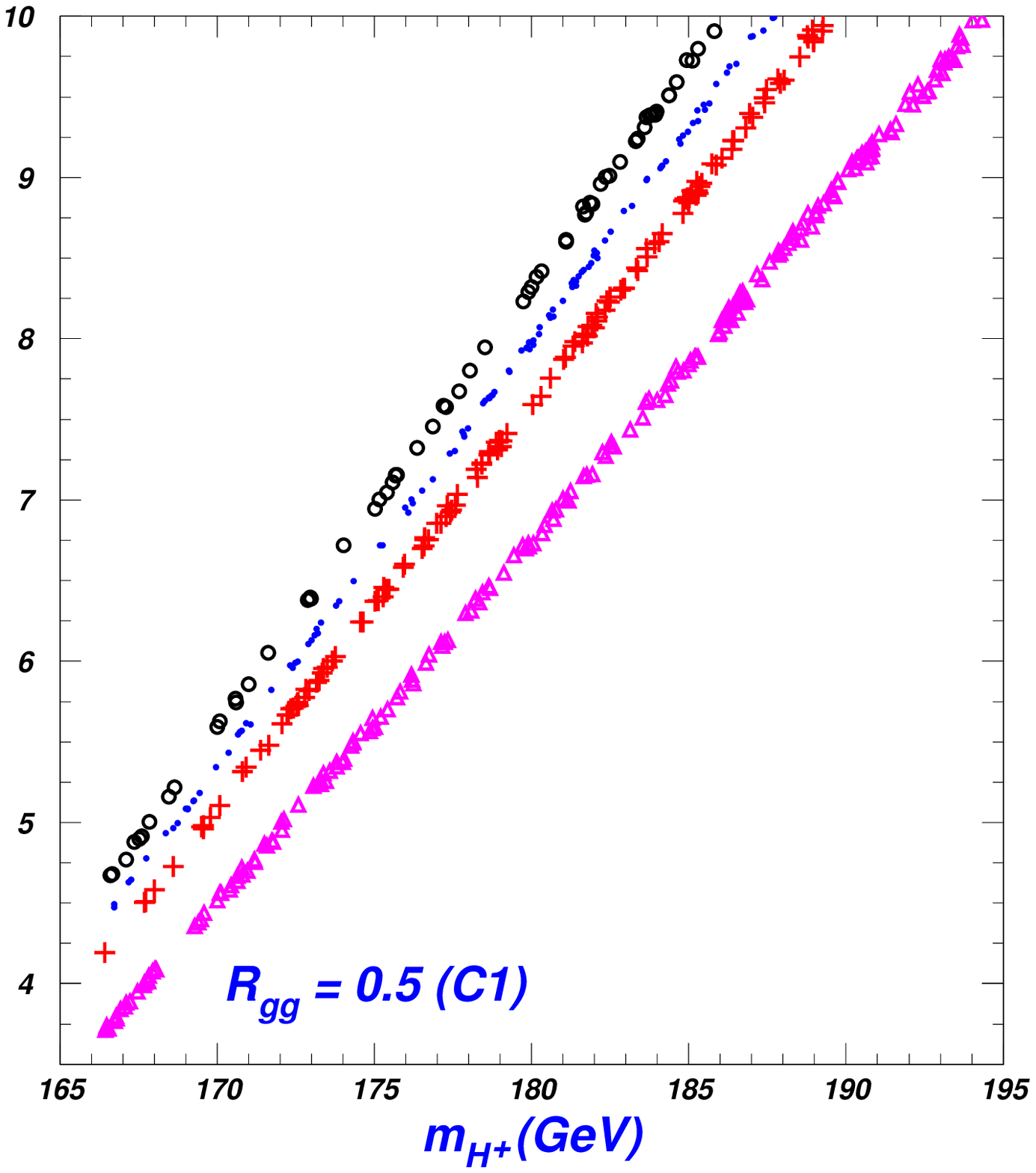,height=6.05cm}
%\end{center}
\vspace{-0.4cm} \caption{For $R_{gg}$ = 0.3 and 0.5, scatter plots
for ($\lambda_1$, $\lambda_4$), ($\lambda_1$, $m_{H^{\pm\pm}}$) and
($\lambda_1$, $m_{H^{\pm}}$). $R_{\gamma\gamma}=2.0$ for triangles
(pink), $R_{\gamma\gamma}=3.0$ for crosses (red),
$R_{\gamma\gamma}=3.5$ for bullets (blue) and $R_{\gamma\gamma}=4.0$
for circles (black), respectively.} \label{rrr35}
\end{figure}

Because $H^{\pm\pm}$ has an electric charge of $\pm 2$, the
$H^{\pm\pm}$ contributions are enhanced by a relative factor 4 in
the amplitude, which can further help $H^{\pm\pm}$ contributions
dominate over the other particle contributions. The sign of the
$H^{\pm}$ and $H^{\pm\pm}$ contributions are respectively determined
by $g_{_{H^{\pm}}}$ and $g_{_{H^{\pm\pm}}}$ which are proportional
to $\lambda_1$ and $\lambda_1+\frac{\lambda_4}{2}$. For $\lambda_1$
and $\lambda_1+\frac{\lambda_4}{2}$ are positive, the $H^{\pm}$ and
$H^{\pm\pm}$ contributions are constructive each other, but
destructively with the contribution of $W$ boson. The masses of
$H^{\pm}$ and $H^{\pm\pm}$ are respectively determined by
$\lambda_4$ from Eq. (\ref{mass}), and vary in the range of 165 GeV
$\sim$ 270 GeV and 110 GeV $\sim$ 320 GeV for the parameter space
taken in Eq. (\ref{space}). Recently, CMS presents the lower bound
of 313 GeV on $H^{\pm\pm}$ mass from the searches for $H^{\pm\pm}$
decaying leptonically \cite{313}. In this model, the limit can be
reduced to 100 GeV since $H^{\pm\pm}$ will also decay into
$W^{\pm}W^{\pm*}$  and $H^{\pm}W^{\pm*}$ \cite{htmrr}. LEP searches for both charged and
neutral scalars give severe constraints on the possible existence of
light scalars \cite{lep}. A conservative lower bound on $m_{H^{\pm}}$ should be
larger than 100 GeV due to the absence of non-SM events at LEP.

The widths of $h$ decay modes at tree-level are the same both in the
HTMVQ and SM. The branching ratio of $h\to \gamma\gamma$ in the
HTMVQ normalized to the SM prediction can be defined as \beq
R_{\gamma\gamma}=\frac{Br(h\to\gamma\gamma)}{Br(h\to\gamma\gamma)^{SM}}
\simeq
\frac{\Gamma(h\to\gamma\gamma)}{\Gamma(h\to\gamma\gamma)^{SM}}.\eeq
The top quark, top partner $t_1$ and $t_2$ contributions depend on
the parameters shown in Table \ref{tabgg}. We take the three points
$B1$, $C1$ and $E1$, and plot $R_{\gamma\gamma}$ versus
$m_{H^{\pm\pm}}$ for $R_{gg}=0.3$, 0.5 and 0.8 in Fig.
\ref{rrrmass}, respectively. The $H^{\pm}$ mass can be determined by
$m_{H^{\pm\pm}}$ from the Eq. (\ref{mass}) and there are the small
mass splitting between $H^{\pm}$ and  $H^{\pm\pm}$. For the small
$m_{H^{\pm\pm}}$, the $H^{\pm\pm}$ and $H^{\pm}$ contributions are
very large and dominant over the other particle contributions, which
leads that $R_{\gamma\gamma}$ reaches $\ord{(10^1)}$ and is not
sensitive to $R_{gg}$. With the increasing of $m_{H^{\pm\pm}}$,
$H^{\pm\pm}$ and $H^{\pm}$ contributions become small and have the
severely destructive interference with other particle contributions,
which leads $R_{\gamma\gamma}$ to be much smaller than 1. Because
the coupling constants of $hH^{\pm\pm}H^{\mp\mp}$ and
$hH^{\pm}H^{\mp}$ increase with $\lambda_1$, the large $\lambda_1$
can enhance sizably the value of $R_{\gamma\gamma}$ for the small
$m_{H^{\pm\pm}}$.

In Fig. \ref{rrr35}, we scan the parameter space shown in Eq.
(\ref{space}), and give ($\lambda_1$, $\lambda_4$), ($\lambda_1$,
$m_{H^{\pm\pm}}$) and ($\lambda_1$, $m_{H^{\pm}}$) for which
$R_{\gamma\gamma}$ equals to 2.0, 3.0, 3.5 and 4.0, respectively. We
only take $R_{gg}$ = 0.3 and 0.5 since $R_{\gamma\gamma}$ is not
sensitive to $R_{gg}$, i.e. the contributions of top quark and top
partners, as long as $R_{\gamma\gamma}$ is much larger than 1. We
stress that $\lambda_4$, $m_{H^{\pm\pm}}$ and $m_{H^{\pm}}$ are
dependent each other according to Eq. (\ref{mass}). Fig. \ref{rrr35}
shows again $R_{\gamma\gamma}$ is not sensitive to $R_{gg}$ when
$R_{\gamma\gamma}$ is much larger than 1. $R_{\gamma\gamma}>2.0$
favors $\lambda_1 >4.0$, $\lambda_4 >0.3$, $m_{H^{\pm\pm}}< 180$ GeV
and $m_{H^{\pm}}< 195$ GeV, where the charged scalars masses can be
in agreement with the current experimental data of LHC and LEP.

\section{The Higgs boson production rates at LHC and Tevatron}

\begin{table}[hbt]
\renewcommand{\arraystretch}{1.5}
\begin{center}
\begin{tabular}{|c|c|ccccc|c|} \hline
\multicolumn{2}{|c|}{}&\multicolumn{5}{c|}{LHC} &  Tevatron\\
\hline $R_{gg}$&$R_{\gamma\gamma}$&~ $h\to\gamma\gamma$~ &~ $h\to
WW^*$ ~&~ $h\to ZZ^*$ ~& $~h\to \tau\bar{\tau}$~ & ~$jjh\to
jj\gamma\gamma$ ~&~$Vb\bar{b}$~\\ \hline $0.2$&$3.5$&~ $1.02$~ &
0.29 & 0.29 & 0.29 & 3.5 & 1.0 \\ \hline $0.3$&$3.0$&~ $1.14$~ &
0.38 & 0.38 & 0.38 & 3.0 & 1.0 \\ \hline $0.3$&$3.5$&~ $1.33$~ &
0.38 & 0.38 & 0.38 & 3.5 & 1.0 \\ \hline $0.5$&$2.0$&~ $1.12$~ &
0.56 & 0.56 & 0.56 & 2.0 & 1.0 \\ \hline $0.5$&$3.0$&~ $1.68$~ &
0.56 & 0.56 & 0.56 & 3.0 & 1.0 \\ \hline $0.5$&$3.5$&~ $1.96$~ &
0.56 & 0.56 & 0.56 & 3.5 & 1.0 \\ \hline $0.6$&$2.0$&~ $1.30$~ &
0.65 & 0.65 & 0.65 & 2.0 & 1.0 \\ \hline $0.6$&$3.0$&~ $1.95$~ &
0.65 & 0.65 & 0.65 & 3.0 & 1.0\\ \hline $0.8$&$2.0$&~ $1.64$~ & 0.82
& 0.82 & 0.82 & 2.0 & 1.0 \\ \hline $0.8$&$3.0$&~ $2.46$~ & 0.82 &
0.82 & 0.82 & 3.0 & 1.0
\\ \hline \hline
\multicolumn{2}{|c|}{CMS}& $1.56^{+0.43}_{-0.43}$ &$0.6^{+0.5}_{-0.4}$& $0.7^{+0.5}_{-0.4}$& $-0.1^{+0.7}_{-1.7}$&$2.1^{+1.4}_{-1.1}$&$\times$\\
\hline
\multicolumn{2}{|c|}{ATLAS}& $1.9^{+0.5}_{-0.5}$ &$0.5^{+0.6}_{-0.6}$& $1.3^{+0.6}_{-0.6}$& $0.5^{+1.5}_{-2.0}$&$\times$&$\times$\\
\hline
\multicolumn{2}{|c|}{CDF-D0}& $\times$ &$\times$& $\times$& $\times$&$\times$&$1.8^{+0.7}_{-0.7}$\\
\hline
\end{tabular}
\renewcommand{\arraystretch}{1.2}
\caption{In HTMVQ, the Higgs boson production rates normalized to
the SM predictions for several values of $R_{gg}$ and
$R_{\gamma\gamma}$. The corresponding measured values at the LHC and
Tevatron are given in the last line.} \label{channel}\end{center}
\end{table}

In Table \ref{channel}, we list the Higgs boson production rates
normalized to the SM predictions for several values of $R_{gg}$ and
$R_{\gamma\gamma}$, and compare them with the corresponding measured
values at the LHC and Tevatron. The measured $jj\gamma\gamma$ rate
at CMS favors $R_{\gamma\gamma}$ in the range of 1.0 and 3.5 since
the Higgs cross section in VBF production are the same both in the
HTMVQ and SM. For $R_{gg}$ = 0.2, the $\gamma\gamma$ rate is
slightly enhanced. For $R_{gg}$ $>$ 0.6, the $\tau\bar{\tau}$ rate
is outside the range of $1\sigma$ of CMS measured value. $R_{gg}$ =
0.5 and $R_{\gamma\gamma}$ = 3.0 can fit the measured Higgs rates at
the LHC and Tevatron relatively well, for which the $\gamma\gamma$
and $WW^*$ rates are respectively between the central values of CMS
and ATLAS. Besides, the Higgs boson in the HTMVQ has two typical
properties: (i) For the signals $VV^*$ ($V=W,~Z$), $\tau\bar{\tau}$
and $b\bar{b}$, the Higgs production rates normalized to their SM
values are the same; (ii) $\sigma(pp\to Vh)\times Br(h\to b\bar{b})$
at the LHC and $\sigma(p\bar{p}\to Vh)\times Br(h\to b\bar{b})$ at
the Tevatron are the same as their SM predictions, respectively.

The LHC diphoton Higgs signal can be well matched in many new
physics models, such as the original HTM \cite{htmrr,htmrr2}, inert
Higgs doublet model (IHDM) \cite{12012644}, Type-II two-Higgs
doublet model (2HDMII) \cite{12071083}, and the model with universal
suppression of Yukawa couplings of fermions (SCFM) \cite{12035083}.
For the HTM and IHDM, the new charged scalars can enhance the decay
width of $h\to\gamma\gamma$ sizably. For the SCFM, the suppressions
of $hb\bar{b}$ and $ht\bar{t}$ couplings can reduce the total Higgs
decay width and Higgs production cross section via gluon-gluon
fusion. For the 2HDMII, the new charged scalars and the modified
Higgs couplings to SM particles can give the corrections to the LHC
diphoton Higgs rate. However, compared to the SM predictions, for
the LHC diphoton Higgs rate is enhanced, the Higgs production rate
into $WW^*$ at the LHC is hardly suppressed for the HTM, IHDM and
2HDMII. The Higgs production rate into $Vb\bar{b}$ at the Tevatron
is always suppressed for the SCFM. These are disfavored by the
recent LHC and Tevatron Higgs data. In the HTMVQ, only the processes
$h\leftrightarrow gg$ and $h\to \gamma\gamma$ at one-loop are
respectively suppressed and enhanced by the virtual contributions of
the vector-like quarks and charged scalars, and the rates for other
processes at tree-level are the same as the SM predictions, which is
favored by the recent LHC and Tevatron Higgs data. Alternatively,
one can introduce the charged and colored scalars or vector bosons
to suppress $h\leftrightarrow gg$ and enhance $h\to \gamma\gamma$,
which is studied in ref. \cite{colorh}.

To get the minimal scenario with only additional vector-like quark
multiplets including $t'$ components able to strongly suppress the
Higgs production via gluon-gluon fusion, an $SU(2)_L$ doublet is
introduced to the HTMVQ, which contains the top partner $t'$ and an
exotic quark with the electric charge 5/3. The exotic quark is
harmless in the parameter fitting since it has no coupling to $h$ at
tree-level. The null results from experimental searches for
fractionally charged heavy baryons or mesons suggest that the
charges of any color triplet quarks should be quantized as
$Q=\frac{2}{3}+integer$ \cite{12073495}, which implies $Q = 5/3$ is
the smallest charge greater than the conventional $2/3$. The exotic
quark with the electric charge 5/3 is also predicted in the model
with $SU(7)$ gauge group \cite{su7} and the model with a left-right
custodial parity invariance of the electroweak symmetry breaking
sector \cite{q53}, respectively.

\section{Conclusion}
In the framework of Higgs triplet model, we introduce some
vector-like quarks in order to explain the recent Higgs boson data
released by LHC and Tevatron collaborations. Compared with the SM,
only the processes $h\leftrightarrow gg$ and $h\to \gamma\gamma$ at
one-loop are modified in this model. The cross section of $gg\to h$
can be sizably suppressed by top partners while $Br(h\to
\gamma\gamma)$ can be more sizably enhanced by the singly and doubly
charged scalars. Therefore, the model will enhance the Higgs
production rates into $\gamma\gamma$ and $jj\gamma\gamma$, and those
for $WW^*$, $ZZ^*$ and $\tau\bar{\tau}$ at the LHC are reduced with
respect to their SM values. The Higgs production rate into
$Vb\bar{b}$ at the Tevatron is the same as the SM value. We find
that the measured Higgs rates at the LHC and Tevatron favor 2.0 $<
R_{\gamma\gamma} <$ 3.5 and disfavor $R_{gg} <$ 0.2 or $R_{gg}>$
0.6. $R_{gg}$ = 0.5 and $R_{\gamma\gamma}$ = 3.0 can fit the
measured Higgs rates at the LHC and Tevatron relatively well, for
which the $\gamma\gamma$ and $WW^*$ rates are respectively between
the central values of CMS and ATLAS.

\section*{Acknowledgment}
We thank Jin Min Yang, Fei Wang and Wen Long Sang for discussions.
This work was supported by the National Natural Science Foundation
of China (NNSFC) under grant Nos. 11005089 and 11105116.

\end{document}